\documentstyle[12pt,epsfig]{article}

\input{epsf}
\newcommand{\bel}[1]{\begin{equation}\label{#1}}
\newcommand{\bal}[1]{\begin{eqnarray}\label{#1}}
\newcommand{\be}{\begin{equation}}
\newcommand{\ee}{\end{equation}}
\newcommand{\ba}{\begin{eqnarray}}
\newcommand{\ea}{\end{eqnarray}}

\newcommand{\bes}{\begin{equation*}}
\newcommand{\ees}{\end{equation*}}
\newcommand{\del}{\partial}

\newcommand {\wt}{\widetilde}

\begin{document}
\begin{titlepage}
\vspace{0.4in}
\begin{center}
{\Large\bf The behavior of the sextic coupling for
the the three-dimensional  $\varphi^4 $ theory}\\
\vspace{.3in} {\large\em Gino N.~J.~A\~na\~nos
\footnote{gananos@ift.unesp.br} and Marcelo P. S.~Pinheiro \footnote{pinheiro@ift.unesp.br}
\\
Instituto de F\'{\i}sica Te\'orica-IFT\\
Universidade Estadual Paulista \\
Rua Pamplona 145, S\~ao Paulo, SP  01405-900  Brazil}
\subsection*{Abstract}
\end{center}
In this work we use the lattice regularization method to study the
behavior of the six point renormalized coupling constant defined at zero
momentum for  the three-dimensional  $\varphi^4 $ theory in the intermediate
and strong coupling domain. The result is in good agreement with the corresponding study
in the Ising limit.
\end{titlepage}
\newpage
\baselineskip .37in
\section {Introduction}
We study in this paper the behavior of the six point renormalized coupling constant
defined at zero momentum for the $(\varphi^4)_{d=3} $ theory for a large variation
of the bare coupling constant in the symmetric phase. For consistent we also show
the behavior of the four point renormalized coupling constant.
In such regime  the perturbation theory doesn't work and the
nonperturbative approach is mandatory. As an analytic alternative
approach in studying this theory in the strong coupling regime it is used the
strong coupling expansion series~\cite{strong1,strong2,strong3}.  The basic idea of this
method consists in factoring out the kinematical parts of the Lagrange from the
path integral and the result of evaluating the remaining non-Gaussian in close form
is a formal expansion of the vacuum functional as a series in inverse
power of coupling constant. It is possible from this expansion  to extract a set of simple
diagrammatic, which can be used to compute the n-point Green's functions of the theory.
For the $(\varphi^4)_{d=3} $ theory it can be shown  from the analytic study
in the strong coupling limit the renormalized
coupling constant  approaches to the asymptotic
limit. The same thing can be said for the six point renormalized coupling
constant. From the point of view of numerical study a relative little work has been done
in order to simulate the general theory in the domain of intermediate to strong coupling
constant. For this reason we were motivated to prepare this paper. This work is a
continuation of previous paper~\cite{phi42d} where the six point renormalized
coupling constant behavior of the $(\varphi^4)_{d=2} $ theory was analyzed. The other important
reason is that $(\varphi^4)_{d=3}$ theory admits a non-trivial continuum limit
~\cite{strong3,Freedman,Freedman2} contrary to $(\varphi^4)_{d=4}$. Much more
effort has been done in the application of renormalization group and high temperature
expansion to the theory $(\varphi^4)_{d=3}$ in the Ising limit\cite{Campostrini:2002cf}.
The techniques that
have been used to determine the proper $n$($n\leq 8$)-point Green functions (1PI) include
high-temperature lattice expansion, Monte Carlo
methods and $\epsilon$ expansion. For the case of $d=3$ a complete list of references
is given in~\cite{Tsypin}. Bender and Boettcher~\cite{Bender}
did the analytic study for renormalized sextic coupling. This study takes into account
the strong coupling calculation on hypercubic lattice in d-dimensions, which were invented
to obtain the continuum limit. In two dimensions
Sokolov and Orlov used renormalization group expansion and Pad\'e-Borel-Leroy resumation
technique to get $g_6$~\cite{Sokolov:1998dq}. In the literature there are a few works
available of the sextic coupling constant using the approach of lattice Monte Carlo.
The first work was performed by Wheater \cite{Wheter} and later in the Ising limit
by Tsypin \cite{Tsypin}. So it seems appropriate the nonperturbative study using the
lattice Monte Carlo technique~\cite{Montvay} to get not only the asymptotic value of the
higher order renormalized coupling constant but also the behavior on the intermediate and
strong regime. Here as usual we consider $ \hbar =c= 1$.
\section{Higher order coupling constants}
We consider in the continuum   the $\phi^4$ theory in $d$-dimensions Euclidean space in the
presence of a source $J$, which  the bare action is given by:
\be S[\phi,J]= \int d^dx
\left[ {1\over 2}(\del_{\mu}\phi)^2+{1\over 2} m^2\phi^2+{g\over4!}\phi^4-J\phi \right].
 \label{action}
\ee
We introduce the vacuum persistent functional $Z[J]$:
\be Z[J]=\int \; D[\phi]\;
\exp(-S[\phi,J])\; \label{zj}
\ee
and from this we define the generating functional $W[J]$ for the
connected Green functions, by writing
\be W[J]=\ln \, Z[J]\, . \label{wj} \ee
The vacuum expectation value of the field $\phi_c$ is given by
\be
\phi_c=<\phi>=\left. \frac{\delta}{\delta J(x)} W[J]\right|_{J=0}
\ee
and the connected $n$-point Greens functions $G_n(x_1 \dots x_n )$
is obtained from the generating functional (\ref{wj}),
\be
G_n(x_1 \dots x_n )=\left.\frac{\delta}{\delta J(x_1)}\dots \frac{\delta}{\delta J(x_n)}
\ln Z(J)\right|_{J=0}\, .
\ee
The effective action  $\Gamma[\phi_c]$ is defined by a functional Legendre
transform of $W[J]$:
\be
\Gamma[\phi_c] =  \int d^dx\; \phi_c(x)\, J(x) - W[J] \; .
\label {legendtr}
\ee
It is well known that the effective action is the generating functional of
one-particle-irreducible(1PI)
vertices, in particular the functional $\Gamma[\phi]$ has a Taylor expansion in powers of
$\phi$ at $\phi=0$;
\be
\Gamma[\phi_c]=\sum_{n} \frac{1}{n}\int d^dx_1 \dots d^dx_n
\;  \Gamma^{(n)}(x_1,\dots,x_n) \, \phi_c(x_1) \dots \phi_c(x_n).
\ee
Here $\Gamma^{(n)}(x_1,\dots,x_n) $ is the proper $n$-point
Green functions (1PI).
Now we consider a source $J$ which is constant and uniform in Euclidean
space-time. This implies that $\phi$ is a constant independent of space-time.
We define the Fourier transform of $\Gamma^{(n)}(x_1,\dots,x_n)$ by
\be
\Gamma^{(n)}(x_1,\dots,x_n)=\int \frac{dk_1}{(2\pi)^d} \dots \int \frac{dk_n}{(2\pi)^d}
\left[ (2\pi)^d\,\delta(\sum_{i=1}^{n}k_i)\; \wt \Gamma^{(n)}(k_1,\dots,k_n) \right]\, .
\ee
Thus, for constant $\phi(x)=\phi$ we have
\be
\Gamma[\phi_c]=\sum_{n=0}^{\infty}\frac{1}{n!}\; \wt \Gamma^{(n)}(0,0,0,\dots,0)
\phi^n \, (2\pi)^d\delta(0)\equiv (2\pi)^d \, \delta(0)\, U(\phi)
\ee
which is the defining equation for the effective potential $U(\phi)$. Thus, the
Taylor coefficients of the effective potential are the 1PI vertices
$\wt \Gamma^{(n)}(0,0,0,\dots,0)$ evaluated at zero external momentum.
The renormalized fourier transform $\wt \Gamma_r^{(n)}(0) $  proper $n$-point
Green functions is obtained as follows:
the wave-function renormalization is obtained from the Fourier transform connected
Green function of two-points from:
\be
Z^{-1}=\left.\frac{d \wt G_2(p^2)}{dp^2}\right|_{p^2=0}
\label{md}
\ee
and the renormalized mass $m^2_r$ is defined by
\be
m^2_r=\left.Z\, \wt G_2^{-1}(p^2)\right|_{p^2=0} \, .
\ee
In general the renormalized $\wt \Gamma_r^{(n)}(0) $  proper $n$-point
Green's functions are given by
\be
\wt \Gamma^{(n)}_r(0) =Z^{n/2}\wt \Gamma^{(n)}(0)  \; ,
\ee
and it follows that the renormalized effective potential can be written as,
\be
U_{r}=\sum_{n=1}^{\infty} \frac{\wt \Gamma_r^{(n)}(0)}{n!} \;\phi_r^{n}
\ee
where $\phi_r=Z^{-1/2} \, \phi$. From here we see that $U_{r}$ is the generating function
of one particle irreducible renormalized Green's function at zero external momentum
on all legs. Since we are doing the study of the theory in the symmetric phase,
we have $\wt \Gamma_r^{(2n+1)}(0)=0$. The particular interest to us is the renormalized
coupling constant $\wt \Gamma^{(4)}_r(0)$ and the renormalized sextic coupling constant
($\wt \Gamma^{(6)}_r(0)$), which can be expressed in terms of Fourier transform of connected
Green functions as follows:
\be
\wt \Gamma_r^{(4)}(0)=\left. -Z^2 (\wt G_2^{-1}(p^2))^4 \wt G_4(p^2) \right|_{p=0}
\label{f4}
\ee
and
\be
\wt \Gamma_r^{(6)}(0)=\left. -Z^3 (\wt G_2^{-1}(p^2))^6 \left( \wt G_6(p^2)-10
\wt G_4^2(p^2) \wt G_2^{-1}(p^2)\right) \right|_{p=0}.
\label{f6}
\ee
The quantities that will be extracted from lattice Monte Carlo simulation
are the dimensionless renormalized zero momentum scattering amplitudes $g_{2n}$ defined by,
\be
g_{2n}=\frac{\wt \Gamma_r^{(2n)}(0)}{m_r^{2n-nd+d}(2n)!}\; .
\ee
\section{Simulation results}
A discrete version of action related to Monte Carlo simulation can be written as
\be S[\phi,J]= \left[ {a^{d-2}\over 2}\sum_{x,\mu}
(\phi_0(x+e_\mu)-\phi_0(x))^2+ {a^d \over 2}\sum_x
m_0^2\phi_0(x)^2+\sum_x{g_0 \over 4!} \phi_0(x)^4 \right] \label {action2}, \ee
where $e_\mu$ is a vector of length $a$ in the positive $\mu$-direction. It's convenient to
work with dimensionless quantities when making numerical simulation. Thus we rescale in
d-dimensions the field, the mass and the coupling constant according to $\phi \equiv
a^{(d/2)-1}\phi_0$, $ m \equiv m_0 a$ and $g\equiv g_0a^{4-d}$  to
make the lattice action independent of the lattice constant $a$.
So we have the action as follows:
\be S[\phi,J]= {1\over 2}\sum_{x,\mu} (\phi(x+e_\mu)-\phi(x))^2+
{1 \over 2}\sum_x m^2\phi(x)^2+\sum_x{g \over 4!}
\phi(x)^4 \label {action2}.
 \ee
As usually we impose periodic boundary condition on fields:
\be
\phi(n+L_{\mu})=\phi(n) \ \ \ \rm {for\  all} \ \ {\mu}\ ,
\ee
We use the standard Metropolis algorithm combined with the Wolff single cluster method
~\cite{Wolff:1989} which is used to avoid the trapping into meta stable states
due to the underlying Ising dynamics. We use the cluster algorithm using the embedded
dynamics for $\phi^4$ theory, according to the action, \cite{Brower:1989mt}
\begin{equation}
\label{Ising} S_{\mathrm{Ising}}  =
\mbox{ } -  \sum_x \sum_{\hat{\mu}}
\left| \phi(x+\hat{\mu}) \phi(x) \right| s(x+\hat{\mu}) s(x) \;,
\end{equation}
where $s(x) = \mathrm{sign}(\phi(x))$. Statistical errors are evaluated taking into account the
autocorrelation time in the statistical sample generated by the Monte Carlo simulation.
For simulation it is used $32^3$ lattice. We choose correlation length
$\xi_r=1/m_r$ as $1<<\xi_r<<L$ to minimize finite size effects. For
numerical simulation purpose it is appropriated to work with the
equivalence Langrange:
\be S[\phi]= 2\kappa \sum_{x,\mu} \phi(x)\phi(x+e_\mu)+
\sum_x \phi(x)^2+\sum_x\lambda
(\phi(x)^2-1)^2 \label {action3},
\ee
where
\be  m^2=\frac{1-2\lambda}{\kappa}-2d \;\;\;\;\;\;\;\;\;\;\;\;\;\;
g=\frac{6\lambda}{\kappa^2}.
\ee
The most part of our effort it is to find the values of $\lambda$ and $\kappa$ for different
values of the bare coupling constants $g$ for a fixed value of the renormalized mass, which
is found within few percent. As we vary the coupling constant to a greater value we have to
put more negative the bare mass $m^2$. Considering the translational invariance of the
correlations functions, one can choose to approximate the momentum derivation in eq.~(\ref{md})
by variation of $\wt G_2(p^2)$ across one lattice spacing and in one direction in order to
calculate the renormalized mass $m_r$
\be
m_r^{2}=\left(\frac{L}{2\pi}\right)^2 \left[ \frac{<\tilde
\phi(0)^2>-<|\tilde \phi(p)|^2>}{<|\tilde \phi(p)|^2>} \right],
\ee
where here $\tilde \phi$ is the Fourier transform of the field and $p=(\frac{2\pi}{L},0)$ is the
smallest available non-zero momentum.

To figure out  $g_4$ and $g_6$ values from simulation we used lattice version of
eqs.(\ref{f4}) and (\ref{f6}),
\be
g_4(4!) =\left. \frac{\wt G_4(p^2)\;\wt G_2^{-2}(p^2)}{\xi_r^d}\right|_{p^2=0}=-\frac{<\wt
\phi(0)^4>-3<\wt \phi(0)^2>^2}{<\wt \phi(0)^2>^2 \xi_r^d},
\label{g4}
\ee
\begin{eqnarray}
g_6 (6!) &=&\left. \frac{10 \wt G_4^2(p^2) \wt G_2^{-4}(p^2)-\wt G_6(p^2)\wt G_2^{-3}(p^2)}
{\xi_r^{2d}} \right|_{p^2=0}
\nonumber \\
&=& 10 g_4^2
-\frac{<\wt \phi(0)^6>-15<\wt \phi(0)^4> <\wt
\phi(0)^2>+ 30<\wt \phi(0)^2>^3}{<\wt \phi(0)^2>^3 \xi_r^{2d}}.
\label{g6}
\end{eqnarray}
The expressions above are prohibited to use in the weak coupling
constant regime due to the large statistical errors. However we
notice that eqs.~(\ref{g4}) and (\ref{g6})  are efficient in the
intermediate and strong coupling regime where the statistical
errors are reasonable ~\cite{Ardekani:1997cy}. We also notice that
the statistical errors increase with value ($L/\xi_r$). The values
in this simulation we find that for the renormalized coupling
constant and sextic coupling constant are consistent with the
prediction of analytic and numerical methods in the Ising limit.
Our prediction for $g_6=2.03 \pm 0.072$ in the Ising limit is in good agreement with value
obtained by Tsypin~\cite{Tsypin},  $g_6=2.05\pm 0.15$.
In figure fig.(1) we present the result of simulations
for a large range of the bare coupling
constant for $g_6$  and $g_4$. In order to be consistent we also include in figure fig.(2)
the result of simulations of $64^2$ lattice in two dimensions which was taken
from previous work~\cite{phi42d}.
We observe both $g_{4}$ and $g_6$, for two and three dimensions,
approach  to asymptotic constant value. This behavior is in
according with numerical and analytical predictions.
From both figures we also can see that the variation of both
quantities are relative small considering the large variation on
$g$. This means the behavior of effective potential in the strong
coupling regime does not change significatively for large
variation of the coupling constant. For the computation of
renormalized higher orders $g_{2n}$ ($n>3$) it is difficult due to
large statistical errors even in the regime of strong coupling .
To avoid this problem it is useful to look at methods, which
calculate the connected Green's functions directly. One of this
method was discussed by Drumond et al~\cite{Drummond,Drummond1}.
They write down for an  action with a source term, the Langevin
equation describing the stochastic evolution of the field in the
theory. Differentiating both sides of the Langevin equation
successively with respect to the source field gives a set of slave
equations, which describe the stochastic evolution of the
estimators of Grenn's function of higher orders~\cite{Weston}.
The interesting study
would be the $(\phi^4)_3$ theory in the broken phase for finite
coupling constant~\cite{Munehisa}.
\begin{figure}[htb]
\centerline{\epsfig{file=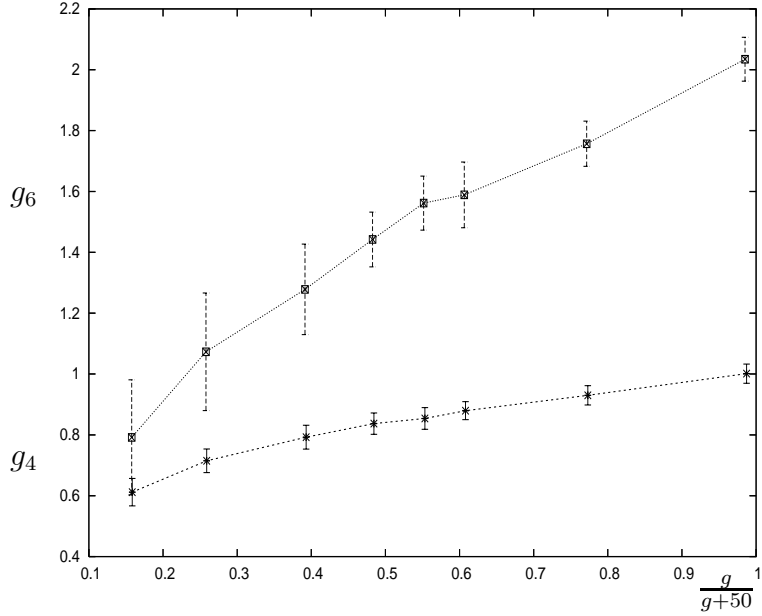,height=10cm,width=8cm,angle=-90}}
\caption{\small{The behavior of $g_4$ and $g_6$ with the bare coupling constant for $d=3$}}
\begin{picture}(120,80)(0,0)
\put(100,150){$g_4$}
\put(100,250){$g_6$}
\put(360,100){\small{$\frac{g}{g+50}$}}
\end{picture}
\end{figure}
\begin{figure}[htb]
\centerline{\epsfig{file=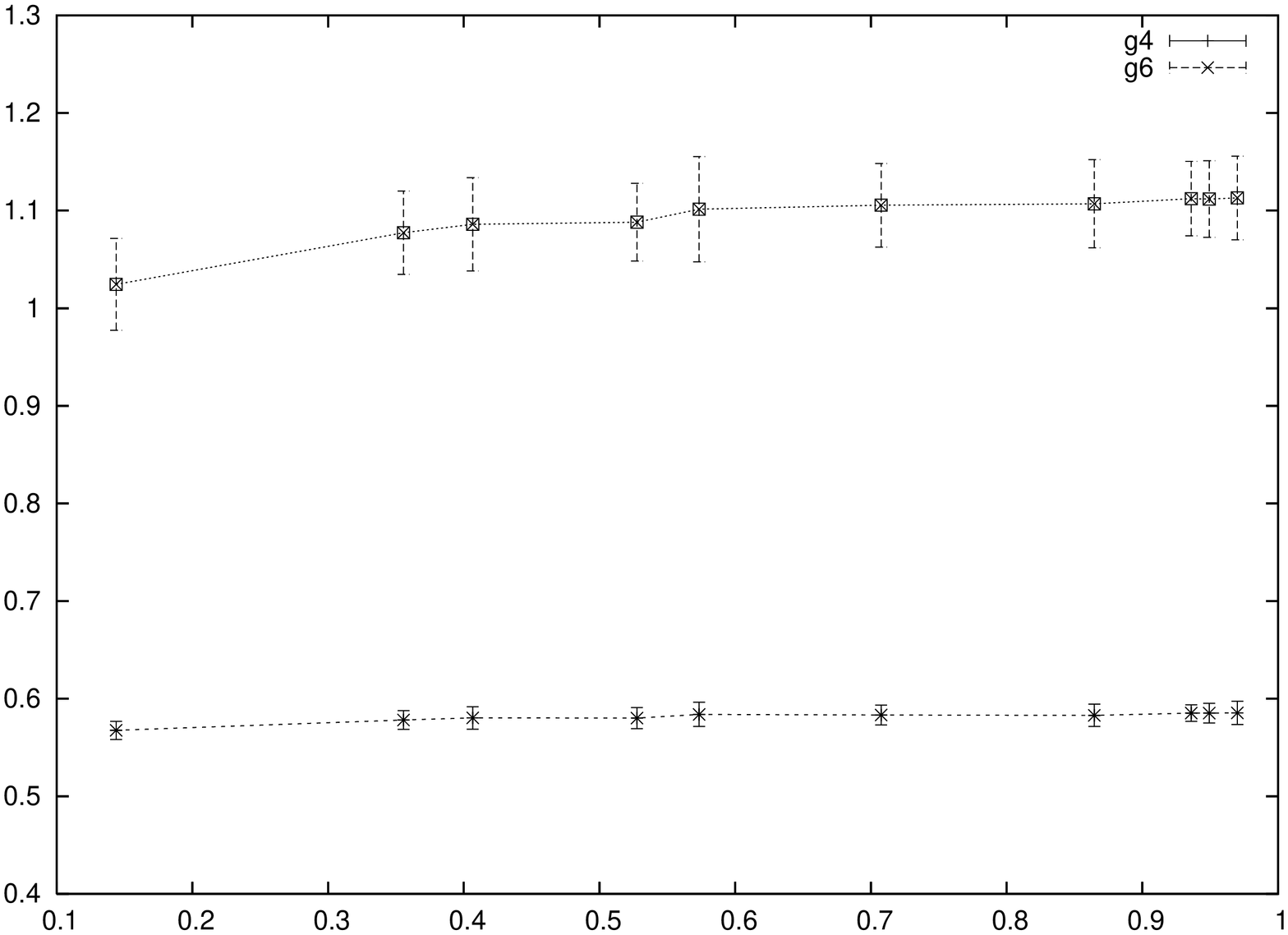,height=10cm,width=8cm,angle=-90}}
\caption{\small{The behavior of $g_4$ and $g_6$ with the bare coupling constant for d=2}}
\begin{picture}(120,80)(0,0)
\put(100,150){$g_4$}
\put(100,250){$g_6$}
\put(360,100){\small{$\frac{g}{g+100}$}}
\end{picture}
\end{figure}

\section{Conclusions}
In this paper we studied on the
lattice the behavior of the renormalized sextic
coupling for the $(\phi^4)_{d=3}$ theory  at intermediate and strong coupling constant domain.
We performed a large variation, from intermediate to
strong coupling, of the bare coupling constant  and notice that the proper six-point
Green functions (1PI) $g_6$ evaluated at zero external momentum has a regular behavior.
This quantity has a definite asymptotic value in the limit of
$g\rightarrow \infty$. Its qualitative behavior is similar to the
two dimensions theory. We remark that in the expansion
 of effective potential in powers of the field the $\phi^6$ term is not negligible.
 Finally our results for $g_4$ and $g_6$ are in good agreement with the values
 available in the literature.
\section{Acknowlegements}
This paper was supported by FAPESP under contract number 03/12271-7


\begin{thebibliography}{10}

\bibitem {strong1} C.~M.~Bender, F.~Cooper, G.~S.~Guralnik and D.~Sharp
Phys.\ Rev.\ D {\bf 19}, 1865 (1979).

\bibitem {strong2} C.~M.~Bender, F.~Cooper, G.~S.~Guralnik and D.~Sharp,
Phys.\ Rev.\ D {\bf 23}, 2976 (1981).

\bibitem {strong3} G.A.~Baker, L.~P.~Benofy, F.~Cooper and D.~Preston,
  Nucl.\ Phys.\ B {\bf 210}, 273 (1982).
\bibitem {phi42d}
G.~N.~J.~Ananos, hep-lat/0512024.
\bibitem {Freedman}
B.~Freedman, P.~Smolensky and D.~Weingarten, Phys. Lett. B {bf 113} 481 (1982).

\bibitem {Freedman2}
F.~Cooper and B.~Freedman, Nucl. Phys. B {\bf 239} 459 (1984).
\bibitem{Campostrini:2002cf}
  M.~Campostrini, A.~Pelissetto, P.~Rossi and E.~Vicari,
  Phys.\ Rev.\ E {\bf 65}, 066127 (2002)
  [arXiv:cond-mat/0201180].
\bibitem{Tsypin}
M.~M.~Tsypin,
Phys.\ Rev.\ Lett.\  {\bf 73}, 2015 (1994).
\bibitem{Bender}
  C.~M.~Bender and S.~Boettcher,
  Phys.\ Rev.\ D {\bf 51}, 1875 (1995)
  [arXiv:hep-th/9405043].

\bibitem{Sokolov:1998dq}
A.~I.~Sokolov and E.V.~Orlov,
Phys.\ Rev.\ B {\bf 58}, 2395 (1998)
[arXiv:cond-mat/9804008].

\bibitem{Wheter}
J.~F.~Wheater,
Phys. Lett.\  {\bf 136B}, 402 (1983).

\bibitem {Montvay} I.\ Montvay, I.\ Munster, {\it Quantum fields on the lattice},
(University of Cambridge Press, 1994).

\bibitem{Wolff:1989}
U. Wolff,
Phys. Rev. Lett. {\bf 62}, 331 (1989)

\bibitem{Brower:1989mt}
R.~C.~Brower and P.~Tamayo,
Phys.\ Rev.\ Lett.\  {\bf 62}, 1087 (1989).

\bibitem{Ardekani:1997cy}
  A.~Ardekani and A.~G.~Williams,
  Phys.\ Rev.\ E {\bf 57}, 6140 (1998)
  [arXiv:hep-lat/9705021].

\bibitem{Drummond}
  I.~T.~Drummond, S.~Duane and R.~R.~Horgan,
  Nucl.\ Phys.\ B {\bf 220}, 119 (1983).

\bibitem{Drummond1}
  I.~T.~Drummond, S.~Duane and R.~R.~Horgan,
  Nucl.\ Phys.\ B {\bf 280}, 25 (1987).

\bibitem{Weston}
  R.~A.~Weston,
  Phys.\ Lett.\ B {\bf 219}, 315 (1989).

\bibitem{Munehisa}
  T.~Munehisa and Y.~Munehisa,
  Z.\ Phys.\ C {\bf 45}, 329 (1989).

\end {thebibliography}
\end{document}